\documentclass[a4paper,onecolumn,11pt, unpublished]{quantumarticle}
\pdfoutput=1
\usepackage[utf8]{inputenc}
\usepackage[english]{babel}
\usepackage[T1]{fontenc}
\usepackage{amsmath}
\usepackage{hyperref}
\usepackage{braket}

\usepackage{tikz}
\usepackage{lipsum}

\usepackage{bbold}

\newtheorem{theorem}{Theorem}
\newtheorem{lemma}{Lemma}[theorem]
\newtheorem{corollary}{Corollary}[theorem]

\newtheorem{definition}{Definition}
\usepackage[a4paper, total={6.2in, 8.3in}]{geometry}

\newcommand{\fid}{\textsf{F}}
\newcommand{\U}{\textsf{U}}
\newcommand{\V}{\textsf{V}}
\usepackage{authblk}
\usepackage[numbers,sort&compress, super]{natbib}
\usepackage{multicol, blindtext}
\usepackage{amssymb}
\usepackage[shortlabels]{enumitem}
\usepackage{qcircuit}
\usepackage[english]{babel}
\usepackage{algorithm}
\usepackage{dsfont}
\usepackage{qcircuit}
\usepackage{color} 
\usepackage[framemethod=TikZ]{mdframed}
\usepackage[framemethod=TikZ]{mdframed}
\usepackage{algorithm}
\usepackage{algpseudocode}
\usepackage{braket}
\usepackage{etoolbox}
\apptocmd{\sloppy}{\hbadness 10000\relax}{}{}
\usepackage[labeled]{multibib}
\newcites{S}{Supplementary Material References}%  \citelatex, \nocitelatex, ...
\usepackage{url}

\usepackage{tikz}
\usepackage{graphicx}
\usetikzlibrary{positioning}

\usepackage[utf8]{inputenc}
\usepackage{pgfplots}
\pgfplotsset{compat=newest}
\usepgfplotslibrary{groupplots}
\usepgfplotslibrary{dateplot}
\usepackage{todonotes}
\usepackage{graphicx}
\usepackage{subcaption}

\usepackage{tocloft}
\begin{document}

\title{Efficient Construction of Quantum Physical Unclonable Functions with Unitary t-designs}

\author{Niraj Kumar}
\email{nkumar@ed.ac.uk}
\affiliation{School of Informatics, University of Edinburgh, EH8 9AB, United Kingdom}
%\orcid{0000-0002-1275-9722}

\author{Rawad Mezher}
\email{rmezher@ed.ac.uk}
\affiliation{School of Informatics, University of Edinburgh, EH8 9AB, United Kingdom}

\author{Elham Kashefi}
\email{ekashefi@inf.ed.ac.uk}
\affiliation{School of Informatics, University of Edinburgh, EH8 9AB, United Kingdom}
\affiliation{CNRS, LIP6, Sorbonne Universit\'{e}, 4 place Jussieu, 75005 Paris, France}

\maketitle
%\tableofcontents

\begin{abstract}
Quantum physical unclonable functions, or \textsf{QPUF}s, are rapidly emerging as theoretical hardware solutions to provide secure cryptographic functionalities such as key-exchange, message authentication, entity identification among others. Recent works have shown that in order to provide provable security of these solutions against any quantum polynomial time adversary, \textsf{QPUF}s are required to be a unitary sampled  uniformly randomly from the Haar measure. This however is known to require an exponential amount of resources. In this work, we propose an efficient construction of these devices using unitary $t$-designs, called $\textsf{QPUF}_t$. Along the way, we modify the existing security definitions of \textsf{QPUF}s to include efficient constructions and showcase that $\textsf{QPUF}_t$ still retains the provable security guarantees against a bounded quantum polynomial adversary with $t$-query access to the device. This also provides the first use case of unitary $t$-design construction for arbitrary $t$, as opposed to  previous applications of $t$-designs where usually a few (relatively low) values of $t$ are known to be useful for performing some task.
We study the noise-resilience of $\textsf{QPUF}_t$ against specific types of noise, unitary noise, and show that some resilience can be achieved particularly when the error rates affecting individual qubits become smaller as the system size increases. To make the noise-resilience more realistic and meaningful, we conclude that some notion of error mitigation or correction should be introduced.
\end{abstract}

\maketitle

\section{Introduction}

%Motivate the need to find good solutions for crypto tasks such as authentication, key distribution etc. 
The need for performing the secure exchange of communication, even in presence of adversaries, has motivated a rich field of cryptography. In its early form, cryptography was mostly concerned with creating ciphertexts that could not be inverted by an enemy eavesdropper. The initial form of cryptography was only based on heuristics and ad-hoc approaches. But with the rise of modern cryptography courtesy Claude Shannon \cite{shannon1948mathematical}, systematic mathematical approaches leading to rigorous security definitions and cryptanalysis has come into prominence. Modern cryptography constructs and analyses protocols against a third party trying to obtain relevant information of the protocol, and provides security in the form of confidentiality, data integrity, authentication, non-repudiation among others \cite{katz2020introduction, goldreich1998modern, mao2003modern}. 

With the advent of internet, information security has become a central talking point and substantial works have gone into proposing cost-effective solutions across distributed networks. An important result in this direction by Canetti and Fishelin states that it is impossible to achieve secure classical cryptographic protocols in practice without making any assumptions on the setup \cite{canetti2001universally}. This result has motivated a new line of research on designing both classical and quantum protocols based on hardware assumptions. An emerging outcome of this is the proposal of quantum physical unclonable function, also called \textsf{QPUF} \cite{arapinis2019quantum, vskoric2012quantum, young2016quantum, nikolopoulos2017continuous, gianfelici2020theoretical, doosti2020client}. These are hardware devices that utilise the properties of quantum mechanics and the random physical disorders that occur in the manufacturing process to construct secure devices. The randomness ensures that the device has the desired security feature of high min-entropy, thus ensuring no dependency on extra cryptographic properties or assumptions. A \textsf{QPUF} is accessed by querying it with a \emph{quantum challenge}, (for example, the quantum state in the form of an electrical signal, an optical pulse, temperature signal, etc.) and obtaining a recognisable \emph{response} that is robust for a particular \textsf{QPUF} token but highly variable for different but very similar \textsf{QPUF}s such that each token seems to output a random response. Each \textsf{QPUF} has a unique identifier associated with it which denotes specific randomisation used in its creation. 

Multiple theoretical and experimental constructions of \textsf{QPUF} have been proposed to date. These include the \emph{quantum read-out PUF}(\textsf{QR-PUF}) where the underlying system is a quantum device but the challenges and responses are classical data. The proposal of \textsf{QR-PUF} included the works of Skoric et. al \cite{vskoric2012quantum}, Nikolopoulos et. al \cite{nikolopoulos2017continuous}, Gianfelici et. al \cite{gianfelici2020theoretical}, and Young et. al \cite{young2016quantum}. The limiting factor in all the above proposals is that they provide the security against only specific bounded adversary attacks such as intercept-resend \cite{vskoric2012quantum}, approximate quantum cloning ~\cite{yao2016quantum}, and challenge estimation attacks ~\cite{vskoric2013security}. This was addressed by the works of Arapinis et. al ~\cite{arapinis2019quantum} who formalised the definition of \textsf{QPUF} as a unitary transformation which satisfies the standard completeness requirements of robustness, uniqueness, and collision-resistance, and thus ensures the crucial security property of unknownness (high-min entropy) which is a primary requirement in building any provably-secure cryptographic protocol. An interesting result that they showed is that in order to prove the unknownness property against any quantum polynomial time (\textsf{QPT}) adversary, the \textsf{QPUF} unitary needs to be sampled uniformly randomly from the Haar measure. The authors however did not prove the crucial uniqueness property - a requirement that allow producing multiple \textsf{QPUF}s sufficiently distinguishable from each other. 

Here, we first address this issue by proving the uniqueness property for unitaries sampled from the Haar measure. This shows that such a construction would indeed be a good candidate construction, as random sampling from the Haar measure produces maximum \emph{single-shot} distinguishable \textsf{QPUF}s with an exponentially (in the device size) high probability. This Haar random sampling construction is however not practically motivated as the resources required in the construction of such a \textsf{QPUF} scales exponentially in the input size ~\cite{knill1995approximation}.

The second part of this work overcomes this exponential requirement in the resources by constructing the quantum physical unclonable functions using approximate unitary $t$-designs \cite{dankert2009exact, brandao2016local}. We refer to this as $\textsf{QPUF}_t$. These unitary designs have the property that they mimic sampling from the Haar measure on the unitary group  up to $t$-th order, but can nevertheless be generated efficiently using quantum circuits of relatively simple structure, and circuit depth scaling polynomially with the input size and the order $t$ of the design. Unitary $t$-designs have found multiple applications including efficient benchmarking of quantum systems via randomised benchmarking \cite{epstein2014investigating}, constructing secure private channels \cite{hayden2004randomizing}, modeling black holes \cite{hayden2007black}, and in providing candidates for devices to exhibit computational speedup using quantum resources \cite{hangleiter2018anticoncentration, harrow2018approximate, bermejo2018architectures}. Our work provides yet another application of these designs in the field of quantum cryptography. Contrary to the previous applications where designs with low $t$ values suffice, we provide, to the best of our knowledge, the first use case of constructing a unitary $t$-design for arbitrary $t$ in a sense that higher the $t$ value, the security of protocols with $\textsf{QPUF}_t$ can be analysed with higher resourced adversaries.

Our candidate construction for $\textsf{QPUF}_t$  is based on  the random quantum  circuit  model (RQC) for generating approximate unitary $t$-designs originally proposed in \cite{BHH16}, but uses at its core a relaxation of technical requirements for RQC recently introduced in \cite{OHS20}. Our construction has a more natural interpretation in the measurement-based model for quantum computing \cite{RB01} (MBQC); as it can be viewed as performing random single qubit $XY$ plane measurements chosen from the interval $[0,2\pi]$ on the non-output qubits of a regularly structured graph state. Equivalently, as in \cite{bermejo2018architectures,mezher2020fault}, our construction can be viewed as a $constant$ $depth$ 2D nearest-neighbour circuit, making its implementation tailor-made on noisy intermediate scale devices \cite{preskill2018quantum}. To avoid any confusion and back and forth translations between circuit model and MBQC, in this work we view our construction only as a RQC, that is, a 1D circuit composed of nearest neighbour two-qubit gates and scaling depth.  %\niraj{In this work we look at our construction in the circuit picture composed of 1D nearest-neighbour interactions.}

We base the security of our $\textsf{QPUF}_t$ construction in the standard game based model as proposed in \cite{arapinis2019quantum}. This involves a manufacturer of the $\textsf{QPUF}_t$ device, clients who intend to use it for their desired task (authentication, key exchange etc.), and a bounded quantum polynomial (\textsf{BQP}) adversary who is allowed limited query access to the device during its transition from the manufacturer to the clients and in between clients. In our work, the security is provided against this adversary and we assume that the manufacturer and the clients are curious but behave honestly. This is a practically motivated setting where the manufacturer intends to establish a secure task with the legitimate clients and wants to prevent impostors from getting valuable secret information of the task. As an example, a smart-phone company could insert the $\textsf{QPUF}_t$ chip with the intention that only the legitimate user can unlock the smart-phone with their fingerprint credential, and no impostor can get into the system without a valid fingerprint.   

In this work, we show that our $\textsf{QPUF}_t$ device construction satisfies the standard security requirements  of robustness, uniqueness and collision resistance. In particular, we provide numerical evidence of our construction exhibiting uniqueness- an essential property to mass produce multiple devices such that they are sufficiently distinguishable to each other.  
Further, we show that a $\textsf{QPUF}_t$ is \emph{practically unknown} to a \textsf{BQP} adversary with limited query access to the device. In order to do so, we modify the existing unknownness criteria introduced by \cite{arapinis2019quantum} to allow for efficient constructions of such devices. This is a crucial result for the security of any protocol built on top of $\textsf{QPUF}_t$ as this enables one to restrict the knowledge that the \textsf{BQP} adversary can gain about the device and thus ensuring that secure protocols can be built on top of it. Another important security issue that one needs to take into consideration is the effect of experimental imperfections the device construction. We show that as long as the imperfection during preparation is restricted to the unitary noise, our noisy device still exhibits all the desired security properties.  

We organise this paper in the following manner. Section~\ref{ing} provides the necessary ingredients required for our work including providing formal definition of a \textsf{QPUF}. We talk about the distance measures and also provide a brief introduction into $t$-designs. Next we revisit the \textsf{QPUF} proposal of Arapinis et. al \cite{arapinis2019quantum} in section~\ref{section:qpufhaar} and prove an essential property of uniqueness which was not proven in the previous work. Section~\ref{sec:qpufdesign} introduces the $\textsf{QPUF}_t$ with an explicit construction and proves the required security properties including robustness, collision resistance and uniqueness. Further, we provide a new \emph{practical} definition of unknownness of \textsf{QPUF} to allow for efficient constructions of these devices. Next, in section~\ref{sec:robust} we allow for imperfections in $\textsf{QPUF}_t$ preparation and show that it is still secure against any \textsf{BQP} adversary with limited query access to the device. Finally, we conclude in section~\ref{conclusion} with some important take away from our work. 

%------------------------Ingredients-----------------------
\section{Ingredients} \label{ing}

This section provides the important tools required construction of $\textsf{QPUF}_t$. We provide a brief formal definition of a $\textsf{QPUF}$ and a brief introduction into unitary $t$-designs. 

\subsection{Quantum Physical Unclonabe Functions} \label{prelim-qpuf}

A quantum PUF, or \textsf{QPUF}, is a secure hardware cryptographic device which utilises the property of quantum mechanics \cite{arapinis2019quantum}. Similar to a classical PUF \cite{armknecht2016towards}, a \textsf{QPUF} is assessed via challenge and response pairs (CRP). However, in contrast to a classical PUF where the CRPs are classical states, the \textsf{QPUF} CRPs are quantum states. 

A \textsf{QPUF} involves a manufacturing process with a quantum generation algorithm, \textsf{QGen}, which takes as an input a security parameter $\lambda$ and generates a PUF with a unique identifier \textbf{id},
\begin{equation}
    \textsf{QPUF}_{\textbf{id}} \leftarrow \textsf{QGen}(\lambda)
\end{equation}
Next we define the mapping provided by $\textsf{QPUF}_{\textbf{id}}$ which takes any input quantum state $\rho_{in} \in \mathcal{H}^{d_{in}}$ to the output state $\rho_{out} \in \mathcal{H}^{d_{out}}$. Here $\mathcal{H}^{d_{in}}$ and $\mathcal{H}^{d_{out}}$ are the input and output Hilbert spaces respectively corresponding to  the mapping that $\textsf{QPUF}_{\textbf{id}}$ provides. This process is captured by the \textsf{QEval} algorithm which takes as an input a unique $\textsf{QPUF}_{\textbf{id}}$ device and the state $\rho_{in}$ and produces the state $\rho_{out}$,
\begin{equation}
    \rho_{out} \leftarrow \textsf{QEval}(\textsf{QPUF}_{\textbf{id}}, \rho_{in})
\end{equation}
It is essential for any cryptographic device to satisfy the correctness and soundness properties. Correctness property ensures that in absence of any adversary, the device validates the honest behaviour with probability close to 1 in the security parameter. Soundness property ensures that the device invalidates the success of any malicious behaviour with probability close to 1 in the security parameter. A \textsf{QPUF} is labelled well constructed if it satisfies these desired properties. The correctness property is characterised by there requirements: robustness, collision-resistance, and uniqueness. 

\noindent 1. \emph{Robustness}: This property is characterised by the indistinguishability of output responses generated by the \textsf{QPUF}. It ensures that if the \textsf{QPUF} is queried separately with two input quantum states $\rho_{in}$ and $\sigma_{in}$ that are $\delta_r$-indistinguishable to each other in the fidelity measure, then the output quantum states $\rho_{out}$ and $\sigma_{out}$ must also be $\delta_r$-indistinguishable,
\begin{equation}
\text{Pr}[\fid(\rho_{out}, \sigma_{out}) \geqslant 1 - \delta_r| \fid(\rho_{in}, \sigma_{in}) \geqslant 1 - \delta_r] \geqslant 1 - \epsilon(\lambda)
\label{eq:robust}
\end{equation}
where $\epsilon(\lambda)$ is a negligible quantity dependent on the desired security parameter. Here $\delta$- indistinguishability for any two quantum states $\rho$ and $\sigma$ is defined as $\fid(\rho, \sigma) \leqslant 1 - \delta$, where $\fid(\rho, \sigma)$ is the fidelity distance measure between the quantum states. Alternatively, other distance measures such as trace norm, euclidean norm (shatten-p norm) can also be used to define security requirements for \textsf{QPUF}. 
    
\noindent 2. \emph{Collision Resistance}: This property is characterised by the distinguishability of output responses generated by the \textsf{QPUF}. It ensures that if the same \textsf{QPUF} is queried separately with two input quantum states $\rho_{in}$ and $\sigma_{in}$ that are $\delta_c$-distinguishable, then the output states $\rho_{out}$ and $\sigma_{out}$ must also be $\delta_c$-distinguishable with an overwhelmingly high probability,
\begin{equation}
\text{Pr}[\fid(\rho_{out}, \sigma_{out}) \leqslant 1 - \delta_c| \fid(\rho_{in}, \sigma_{in}) \leqslant 1 - \delta_c] \geqslant 1 - \epsilon(\lambda)
\label{eq:collision}
\end{equation}
\noindent 3. \emph{Uniqueness}: This property captures the \textsf{QPUF} generation process to ensure that sufficiently distinguishable \textsf{QPUF}s are generated. This is captured by mapping the each \textsf{QPUF} as a quantum operation characterised by a completely positive trace preserving (CPTP) map that takes the input quantum states in $\mathcal{H}^{d_{in}}$ to output states in $\mathcal{H}^{d_{out}}$. We say that two such maps $\textsf{QPUF}_{\textbf{id}_i}$ and $\textsf{QPUF}_{\textbf{id}_j}$ are $\delta_u$ distinguishable if,
\begin{equation}
\text{Pr}[ \| \textsf{QPUF}_{\textbf{id}_i} - \textsf{QPUF}_{\textbf{id}_j} \|_\diamond \geqslant \delta_u | i\neq j] \geqslant 1 - \epsilon(\lambda) 
\end{equation}
where $\|.\|_{\diamond}$ is the diamond norm distance measure for the distinguishablity between any two \textsf{QPUF}s.

The parameters $\delta_r, \delta_c$ and $\delta_u$ are determined by the security parameter $\lambda$. For \textsf{QPUF}, the two constraints $\delta_r \leqslant \delta_u$ and $\delta_r \leqslant \delta_r$ ensure a clear distinction between different \textsf{QPUF}s.

Once the requirements of a well constructed \textsf{QPUF} is defined, one often looks for constructions that satisfy these requirements. It was shown by Arapinis et. al \cite{arapinis2019quantum} that a unitary device sampled from Haar measure set of unitaries satisfies the robustness and collision resistance requirements of a \textsf{QPUF}. In this work, we show that such a construction also satisfies the uniqueness property, thus establishing that it is a suitable candidate for a \textsf{QPUF}.  This however is not a practical construction as it is known to require exponential resources in the input size. Hence our work focuses on an alternative resource efficient construction of \textsf{QPUF} based on the random circuit model for generating approximate unitary $t$-designs~\cite{BHH16}. In Section~\ref{sec:qpufdesign}, we show that this construction satisfies all the \textsf{QPUF} requirements.

 Next, in order to build cryptographic applications such as authentication, device identification among others on top of \textsf{QPUF}, one requires that they satisfy the soundness property. This is characterised by `unknownness'.  Also referred to as min-entropy property or randomness property of \textsf{QPUF}, it is quantified by the amount of knowledge possessed by the adversary about the \textsf{QPUF} before the start of a \textsf{QPUF}-based protocol. We provide a formal definition of unknownness in the Sections~\ref{section:qpufhaar} and \ref{sec:qpufdesign}. This unknownness property is highly crucial in restricting the pre-protocol knowledge of the adversary and to prove that the \textsf{QPUF} construction is `selectively unforgeable' which is a requirement to prove security in applications built on top of \textsf{QPUF}. Formally the device is selectively unforgeable against a quantum polynomial time (\textsf{QPT}) adversary if given access to a polynomial number of challenge-response pairs of \textsf{QPUF}, the probability that the adversary receives a new random challenge $\rho_{\text{in}}$ from a set of challenges \textsf{Set}, and preforms a local operation $\mathcal{A}_{\textsf{QPT}}$ to produce the response $\sigma \leftarrow \mathcal{A}_{\text{QPT}}(\rho; S \in (\rho_i,\sigma_i)_{1\leqslant i \leqslant \text{poly}(n)} )$ which is $\mu$-indistinguishable from the response produced by \textsf{QPUF}, $\rho_{\text{out}}$ on the same challenge, is bounded by a factor $\epsilon$, 
    \begin{equation}
    \begin{split}
        \textsf{Pr}[\fid(\sigma, \rho_{\text{out}}) \geqslant 1 - \mu|\rho \stackrel{\$}{\leftarrow} \textsf{Set}, S ] \leqslant \epsilon(\lambda)
        \end{split}
    \end{equation}

\subsection{Distance Measure}

\noindent 1. \emph{Fidelity}. This is a measure of distance between two quantum density operators. Although, not being a metric itself, it is related to the metric, \emph{Bures distance} $\textsf{D}_b = \sqrt{2 - 2\fid}$. The fidelity of density operators $\rho$ and $\sigma$ is defined as,
\begin{equation}
\fid(\rho, \sigma) = \text{Tr}\big[\sqrt{\sqrt{\rho}\sigma\sqrt{\rho}}\big]
\end{equation}
\noindent 2. \emph{Diamond norm}. Diamond norm is a distance metric for any two completely positive trace preserving quantum operations $\U_1$, $\U_2$. It is defined as,
\begin{equation}
\| \U_1 - \U_2 \|_{\diamond} = \underset{\rho}{\text{max}}(\|(\U_1 \otimes \mathbb{I})[\rho] - (\U_2 \otimes \mathbb{I})[\rho] \|_1)
\end{equation}
Operationally it quantifies the maximum probability of distinguishing operation $\U_1$ from $\U_2$ in a single use. More generally, the diamond norm of an operator $A$  has this form defined as, 
$$\|A\|_{\diamond}=sup_l\|A \otimes \mathbb{I}_{l}\|_{1,1},$$
where $$\|B\|_{1,1}=sup_{X \neq 0}\frac{\|B(X)\|_1}{\|X\|_1},$$
and $\|.\|_1$ is the usual trace  norm.

\subsection{Haar Measure Group}

A Haar measure is a non-zero measure on any locally compact group $G$ such that $\mu : G \rightarrow [0, \infty]$ such that for all $X \subseteq G$ and $x \in G$:
\begin{equation}
    \mu(xX) = \mu(Xx) = \mu(X)
\end{equation}
where,
\begin{equation}
    \mu(X) = \int_{x \in G} d\mu(x)
\end{equation}

In particular, the Haar measure $\mathsf{d}\mu(\U)$ can be defined for a unitary group $\U(d)$ acting on $\log_2 d$ qubits. Choosing unitaries from the Haar measure on $\U(d)$ can be thought as a natural notion of uniform sampling from the unitary group. In practice however, sampling from the Haar measure is exponentially costly \cite{knill1995approximation}.   As an example, for $d=2$, the unitary matrix can be represented as $\U = e^{i\phi}$, with $0\leqslant \phi \leqslant 2\pi$. The Haar measure $\mathsf{d}\mu(\U) = \mathsf{d}\phi$ can then be easily verified to measure the uniformity of the unitary group, since $\mathsf{d}\mu(\U_0\U) = \mathsf{d}\mu(\U\U_0) = \mathsf{d}\mu(\U)$, which translates to
$\mathsf{d}(\phi + \phi_0) = \mathsf{d}(\phi)$ for any fixed unitary $\U_0 = e^{i\phi_0}$,
\begin{equation}
    \mu(\U) = \int_{x \in \U} \mathsf{d}\mu(x) = \int_{0}^{2\pi} \mathsf{d}\phi
\end{equation}

For a general $d$, the Haar measure on the unitary group can be similarly constructed by invoking the group invariant measure $\mathsf{d}\mu(\U_0\U) = \mathsf{d}\mu(\U\U_0) = \mathsf{d}\mu(\U)$ for any fixed $\U_0 \in \U(d)$.

\subsection{Unitary $t$-designs} \label{sec:prel-design}

A unitary $t$-design on $\U(d)$ is a set of $\log_2 d$-qubit unitaries $\mathcal{U}=\{\U_i\}_{i=1,..,k} \subset \U(d)$ together with a probability distribution  over $\mathcal{U}$ which results in sampling each unitary $\U_i \in \mathcal{U}$ with probability $p_i \geq 0$ ($\sum_{i=1,\cdots,k}p_i=1$) for $i=1,\cdots,k$. We will denote a unitary $t$-design as a set of couples $\{p_i,\U_i\}_{i=1,\cdots,k}$. A unitary $t$-design \emph{mimics} sampling from the Haar measure up to $t$-th order in the statistical moments either \emph{exactly} (called exact unitary $t$-designs) or \emph{approximately} up to some precision $\varepsilon$ (called $\varepsilon$-approximate $t$-designs) \cite{DCE+09}. The construction of exact unitary $t$-designs for any $t$ and any dimension $d$ of the unitary group has long been a notoriously difficult task \cite{B19}. Recently however, \cite{BN20} showed how to construct exact unitary $t$-designs for any $t$ and $d$; although the quantum circuits needed to implement these constructions were not discussed in \cite{BN20}. Since we are interested in practical implementations of \textsf{QPUF}s achievable by means of quantum circuits of well-known and relatively simple structure, we will focus in this paper on sampling from $\varepsilon$-approximate $t$-designs, a task which has been shown possible by using the simple model of random quantum circuits (RQC) \cite{BHH16}.

Now we go on to defining an $\varepsilon$-approximate $t$-design. Various definitions of approximate  designs suitable for various applications have been proposed (see for example \cite{HM18} for an overview of these definitions), the definition we will focus on henceforth in this paper is that of an $additive$ $\varepsilon$-approximate $t$-design, since this is sufficient for our purposes. Thus, whenever we say $\varepsilon$-approximate $t$-designs throughout this paper we will mean additive $\varepsilon$-approximate $t$-designs, unless specified otherwise. Let $\mu_H$ be the Haar measure on $\U(d)$, $\{p_i,\U_i\}_{i=1,\cdots,k}$ is said to be an (additive) $\varepsilon$-approximate unitary $t$-design if the following holds
$$\|\delta_t-\delta_t^{H}\|_{\diamond} \leq \varepsilon.$$
Here, $\delta_{t}$ is a quantum channel defined as
$$\delta_{t} (Y)=\sum_{i=1,\cdots,k}p_i\U_i^{\otimes t}Y\U_i^{\dagger \otimes t},$$ where $Y$ is  a  density matrix on a Hilbert space of dimension $d^{t}$. The equivalent Haar random quantum channel associated with $Y$ is, $$\delta^{H}_{t} (Y)=\int \U^{\otimes t}Y \U ^{\dagger \otimes t} \mu_H(d\U).$$ 
Note that any approximate $t$-design is also an approximate $m \leq t$-design. Using this we construct the $\textsf{QPUF}_t$ in section~\ref{sec:qpufdesign}.

\section{\textsf{QPUF} construction from Haar random Unitary} \label{section:qpufhaar}

\begin{figure}[h!]
\includegraphics[scale=0.45]{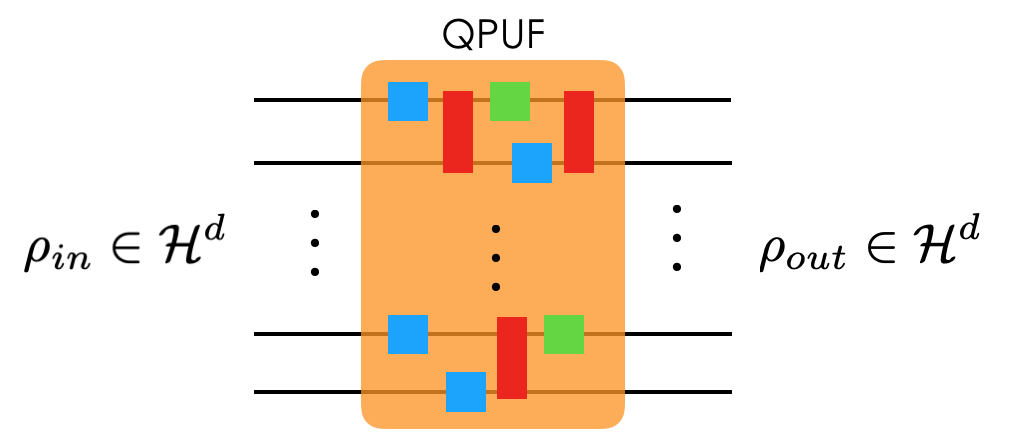}
\centering
\caption{Illustration of \textsf{QPUF} as a unitary operation with input and output quantum states in $\mathcal{H}^d$. The blue and green boxes are single-qubit gates, while red boxes are two-qubit gates. These are the building blocks for the qPUF construction.}
\label{fig:qpuf}
\end{figure}

In this section, we revisit the \textsf{QPUF} construction proposed by \cite{arapinis2019quantum}. The authors showed that the desired correctness requirements of robustness and collision-resistance of \textsf{QPUF} of size $d$ can be satisfied by a unitary map of dimension $d$ sampled from the Haar measure unitary set. Further they also showed that it satisfies the crucial unknownness property against any \textsf{QPT} adversary.
Figure~\ref{fig:qpuf} provides an illustration of such a unitary map which takes an input state $\rho_{in} \in \mathcal{H}^{d}$ and maps to an output state $\rho_{out} \in \mathcal{H}^{d}$. Our contribution here is to explicitly show that this construction satisfies the crucial uniqueness property too which is crucial to `mass-produce' sufficiently distinguishable \textsf{QPUF}s. 

\subsection{Robustness $\&$ Collision-Resistance} \label{sec:robust-haar}

It is fairly straighforward to check that any unitary map satisfies the robustness and collision-resistance requirements. This is due to the crucial property of a unitary map $\U$ that $\U^{\dagger}\U = \mathbb{I}$. Any unitary acting independently on two input states $\rho_{in}$ and $\sigma_{in}$, produces the corresponding output state $\rho_{out} = \U\rho_{in}\U^{\dagger}$ and $\sigma_{out} = \U\sigma_{in}\U^{\dagger}$ respectively. Since the fidelity is a unitary invariant measure, hence $\fid(\rho_{in}, \sigma_{in}) = \fid(\rho_{out}, \sigma_{out})$. Hence this implies the both robustness and collision-resistance properties in a `strict' sense i.e. Eq~\ref{eq:robust} and \ref{eq:collision} are satisfied with a probability 1.

\subsection{Uniqueness}

The previous two requirements can be satisfied by any unitary map. However, in order to produce multiple sufficiently distinguishable \textsf{QPUF}s, we show that sampling a \textsf{QPUF}s from a Haar random unitary set is sufficient to ensure maximum distinguishability with exponentially high probability. We note that, this is not the only method to produce \textsf{QPUF}s which satisfy the uniqueness property. One could alternatively ensure that they pick distinguishable unitaries in the manufacturing process enforce the uniqueness requirement. However, this method would incur two key problems. One being that in order to ensure that the two unitaries picked are sufficiently distinguishable, the manufacturer of the \textsf{QPUF} needs to know the full description of the unitaries. This information would not normally be known to the manufacturer especially if there are random purtubations (in the form of noise) in the manufacturing process. Secondly, as we see in the next section, this would compromise the unknownness property, a key requirement to prove security of applications built on top of \textsf{QPUF}. 

Next, for the uniqueness property we use the following theorem,

\begin{theorem}
If \textsf{QGen} algorithm samples two unitaries $\textsf{QPUF}_{\text{id}_i}$ and $\textsf{QPUF}_{\text{id}_j}$ uniformly randomly from the Haar measure set of unitaries, then the probability that they are $\delta_u$-far is exponentially close to 1. 
\end{theorem}

\noindent \emph{Proof}: Let us denote the two \textsf{QPUF}s sampled uniformly randomly from the Haar measure set of untiaries to be $\textsf{QPUF}_{\text{id}_i} = \textsf{U}_0$ and $\textsf{QPUF}_{\text{id}_j} = \textsf{U}_1$. Then for any positive operator $X$, we can construct two unitary channels given by  $\Phi_0(X) = \textsf{U}_0 X \textsf{U}_0^{\dagger}$ and $\Phi_1(X) = \textsf{U}_1 X \textsf{U}_1^{\dagger}$. For these two channels, the diamond norm can be expressed as,
\begin{equation} \label{diamond_numerical_range}
\begin{split}
    \| \Phi_0 - \Phi_1 \|_{\diamond} & =  \underset{\rho}{\text{max}}(\|(\Phi_0 \otimes \mathbb{I})[\rho] - (\Phi_1 \otimes \mathbb{I})[\rho] \|_1) \\
    & = 2 \sqrt{1 - \delta (\textsf{U}_0^{\dagger}\textsf{U}_1)^2}
\end{split}
\end{equation}

where $\delta(X)$ is the minimum absolute value taken over the numerical range of the operator $X$ i.e.,
\begin{equation}
\delta(X) = \underset{\ket{\phi}}{\text{min}}|\bra{\phi}X\ket{\phi}|
\end{equation}
where the minimum is over all the input states $\ket{\phi}$.
Since the diamond norm is unitarily invariant, hence with no loss of generality, one can set $\U_0 = \mathbb{I}$ and $\U_1 = \U$ where $\U$ is picked uniformly randomly from the Haar measure.
Let us denote the numerical range of a Haar random unitary $\U$ as $W(\U)$,
\begin{equation}
    W(\U) = \{ \bra{\phi}\U\ket{\phi} : \braket{\phi|\phi} = 1 \}
    \label{Eq:numericalrange}
\end{equation}
It can be easily checked that the numerical range is unitarily invariant, i.e. $W(\textsf{V}^{\dagger}\U\textsf{V}) = W(\U)$ for any unitary matrix \textsf{V}. Since a unitary matrix is also a normal matrix, hence we can write a unitary $\U = \textsf{V}^{\dagger}\Lambda\textsf{V}$, where $\Lambda := \text{diag}(\lambda_i)_{i=1}^{d}$, where $\Lambda$ has eigenvalues of \U on its diagonal and \textsf{V} is a unitary matrix. 

Let us denote a unit quantum state vector $\ket{\phi} := [\phi_i, \cdots, \phi_d]^T$. From Eq~\ref{Eq:numericalrange} and the unitarily invariant property of numerical range, we obtain
\begin{equation}
    W(\U) = W(\Lambda) = \bra{\phi}\Lambda\ket{\phi} = \sum_{i=1}^{d} \lambda_i |\phi_i|^2
\end{equation}
where $|\phi_{i}| \geqslant 0$ for all $i \in d$ and $\sum_{i=1}|\phi_i|^2 = 1$. Thus the numerical range of a unitary matrix $\U$ is just a convex hull of its eigenvalues. 

So the problem of obtaining a lower bound on the diamond norm reduces to bounding the absolute value of the numerical range of $\U$:
\begin{equation}
\begin{split}
    &\delta(\U) := \underset{t_1,\cdots,t_d}{\text{minimize}} \hspace{2mm} |\sum_{i=1}^{d} \lambda_i t_i| \\
    &\text{subject to} \hspace{2mm} t_i \geqslant 0 \hspace{2mm}\forall i \in [d], \sum_{i=1}t_i = 1
\end{split}
\end{equation}

First let us consider the eigenvalues of a general unitary matrix $\U$. These eigenvalues are in general complex numbers with absolute value 1. Let us denote them by $\{e^{i\alpha_1},\cdots,e^{i\alpha_d}\}$. To prove the main theorem, we first prove the following lemma.

\begin{lemma}  \label{lemma:num-range-theta}
If all the eigenvalues $\{e^{i\alpha_1},\cdots,e^{i\alpha_d}\}$ of a general unitary matrix $\U$ lie in an arc of size $\theta$, then $\delta(\U)$ is lower bounded by a function of $\theta$. 
\end{lemma}

\noindent \emph{Proof}: If all the eigenvalues lie in the arc $\theta \leqslant \pi$ then there exists $j_0,k_0$ such that $|\alpha_{j_0} - \alpha_{k_0}| = \theta$. For all other $j,k \in [D]$, $|\alpha_j - \alpha_k| \leqslant \theta$. If the arc $\theta > \pi$ then there exists $j_0, k_0$ such that $|\alpha_{j_0} - \alpha_{k_0}| = \pi - \underset{j,k \in [D]}{\text{min}}[\pi - |\alpha_j - \alpha_k|]$. 

To find $\delta(\U)$, we first bound $|\sum_{j=1}^{d}e^{i\alpha_j}t_j|$,
\begin{equation}
\begin{split}
\hspace{2mm} |\sum_{j=1}^{d}e^{i\alpha_j}t_j| 
&= \Big[(\sum_{j=1}^{d} t_j \cos\theta_j)^2 + (\sum_{j=1}^{d} t_j \sin\theta_j)^2\Big]^{1/2} \\
&= \Big[(\sum_{j=1}^{d} t_j^2  + 2\sum_{j \neq k}t_jt_k\cos(\theta_j - \theta_k)\Big]^{1/2} \\
&= \Big[1 - 2\sum_{j\neq k}t_jt_k[1 - \cos(\theta_j - \theta_k)]\Big]^{1/2} \\
&\geqslant \Big[ 1 - 2t_{j_0}t_{k_0}[1 - \cos(\theta_{j_0} - \theta_{k_0})]\Big]^{1/2} \\
\end{split}
\label{Eq:bound}
\end{equation}
where, from 3-rd to 4-th line, we have used the convexity argument that given $t_j \geqslant 0$ for all $j \in [D]$ and $\sum_{j=1}^{D}t_j = 1$, we get the maximization of $\sum_{j\neq k}t_jt_k[1 - \cos(\theta_j - \theta_k)]$ when $t_i \neq 0$ for $j_0, k_0$ and is 0 otherwise. This implies that even if the arc is spanned by multiple eigenvalues, the maximisation occurs with a convex span of only two eigenvalues $\{e^{i\alpha_{j_0}}, e^{i\alpha_{k_0}}\}$.

Now when the convex span is over only two eigenvalues, then $t_{j_0} = 1 - t_{k_0}$. From Eq~\ref{Eq:bound}, we can compute the quantity $\delta(\U)$ as,
\begin{equation} \label{numerical-range-theta}
\begin{split}
    \delta(\U) &= \underset{t_{j_0}}{\text{min}} \hspace{2mm} \sqrt{1 - 2t_{j_0}(1 - t_{j_0})[1 - \cos(\theta_{j_0} - \theta_{k_0})]} \\
    &= \sqrt{\frac{1}{2} + \frac{1}{2}\cos(\theta_{j_0} - \theta_{k_0})}
\end{split}
\end{equation}
This completes the proof of the lemma. Few corollaries that emerge as a consequence of this lemma are,

\begin{corollary}\label{c1}
For $\theta \leqslant \pi$, the minimum of absolute value of number range of a unitary $\U \in \U_{D}$ is $\delta(\U) = \sqrt{\frac{1}{2} + \frac{1}{2}\cos\theta}$
\end{corollary}
\begin{corollary}\label{c2}
If $\theta > \pi$, then for large $d$ and when the eigenvalues are normally distributed along any specified sub-length of the arc, the minimum of absolute value of number range of a unitary $\U$ is $\delta(\U) \approx 0$. This is because there exists $j_0, k_0$ such that $|\alpha_{j_0} - \alpha_{k_0}| = \pi - \underset{j,k \in [D]}{\text{min}}[\pi - |\alpha_j - \alpha_k|] \approx \pi$ in the limit of large $d$.
\end{corollary}

Now to prove the main theorem, we first use the result from \cite{wieand2002eigenvalue} which states the following: Suppose a unitary $\U \in \U(d)$ is sampled randomly from the Haar measure. Then for a fixed arc $\theta$ of a unit circle, let $Y$ be the random variable corresponding to the number of eigenvalues of $\U$ which lie in the $\theta$ arc. Then as the size $d$ grows, the quantity $Z := (Y - \mathbb{E}[Y])/\sqrt{\sigma(Y)}$ is asymptotically normally distributed, where $\sigma(Y)$ is the variance of $Y$.

Suppose we look at such a random variable $Y$ for the fixed arc $\theta = \pi - \epsilon$, with $\epsilon \ll 0$. Then \cite{wieand2002eigenvalue} show that, the expected value of $Y$ is,
\begin{equation}
    \mathbb{E}(Y) = \frac{d(\pi - \epsilon)}{2\pi} \approx \frac{d}{2}
\end{equation}

and the variance,
\begin{equation}
\begin{split}
    \sigma(Y) &= \frac{1}{\pi^2}(\log d + 1 + \gamma + \log \left|2\sin(\frac{\pi-\epsilon}{2})\right|) + o(1) \\
    &\approx \frac{1}{\pi^2}(\log d + 1 + \gamma + \log 2) + o(1)
\end{split}
\label{var}
\end{equation}
where $\gamma \approx 0.57721..$ is Euler's constant.

Thus it means the for $\theta \approx \pi$, on expectation, about half the number of eigenvalues lie within the arc, with the variance given by Eq~$\ref{var}$. Also in the large $d$ limit, since $Z$ follows a normal distribution, then it follows using Chernoff-Hoeffding bound that,
\begin{equation}
    \text{Pr}[|Y - \mathbb{E}[Y]| > \mathbb{E}[Y]] \leqslant e^{-\mathbb{E}[Y]^2/2\sigma(Y)^2} \approx e^{-d\pi^2/4\log d}
\end{equation}

This result implies that when any two $\textsf{QPUF}_{\textbf{id}_i}$ and $\textsf{QPUF}_{\textbf{id}_j}$ are sampled uniformly randomly from the Haar measure, then probability that they are not perfectly distinguishable i.e. $\| \textsf{QPUF}_{\textbf{id}_i} - \textsf{QPUF}_{\textbf{id}_j} \|_\diamond \leqslant 2$ is exponentially low.  Next we briefly mention the required unknownness property of the \textsf{QPUF}.

\subsection{Unknownness property} \label{sec:haar-qpuf-unknown}

In this section, we revisit the `unknownness' definition provided by \cite{arapinis2019quantum} and highlight the drawback of this definition in allowing for resource efficient \textsf{QPUF} constructions. The original  unknown definition is captured in the following definition (slightly modified  \cite{arapinis2019quantum}),
\begin{definition}[\textsf{$\epsilon,\lambda,\delta,$-Unknown \textsf{QPUF}}] \label{def:unknownhaar}
We say that a \textsf{QPUF} constructed as a unitary transformation $\U$ from a  set $S \subseteq \U(d)$ is $(\epsilon,\lambda,\delta)$-unknown, if for any  quantum polynomial time (QPT) adversary $\mathcal{A}$, before making any query to $\U$, the probability of outputting a  response $\mathcal{A}(\rho)$ with fidelity at least $1-\delta$ with respect to the \emph{ideal} response $U\rho U^{\dagger}$  on every state $\rho=|\psi\rangle \langle \psi|$, with $|\psi\rangle \in \mathcal{H}^d$ is bounded by:
\begin{equation}
\mathsf{Pr}[\forall |\psi\rangle \in \mathcal{H}^d: \mathsf{F}(\mathcal{A}(\rho),\U\rho \U^{\dagger}) \geqslant 1 - \delta] \leqslant \epsilon(\lambda)
    %\underset{\U\leftarrow S}{\mathsf{Pr}}[\mathsf{F}(\mathcal{A}(\rho),\U\rho \U^{\dagger}) \geqslant 1 - \delta] \leqslant \epsilon(\lambda)
    \label{eq:unknownhaar}
\end{equation}
\end{definition}

An immediate consequence of this definition is that if the $U$ is a Haar random unitary, then $\epsilon(\lambda) = \textsf{negl}(log_2(d))$ for any $\delta$ with $1-\delta=\textsf{non-negl}(log_2(d))$ \cite{arapinis2019quantum}. Thus this definition ensures that apriori, the adversary has negligibly small information of the \textsf{QPUF}, and thus it is essentially unknown. 

The definition~\ref{def:unknownhaar} bounds the prior knowledge of any QPT adversary $\mathcal{A}$ in terms of their ability to correctly produce the output states given any random state $\rho$ picked from the Haar measure set of states. However, one major drawback of this definition is that it limits the capability of `any' QPT adversary. This automatically rules out a \textsf{QPUF} construction which is composed of polynomial sized circuit since if there was such a \textsf{QPUF} construction, then there would exist `a' QPT adversary which would generate the correct response of any state $\rho$ with a probability 1. In that case, that adversary would be described by the \textsf{QPUF} unitary itself. 

In the next section, we propose a construction of \textsf{QPUF} using resource efficient unitary $t$-designs i.e. the description of the unitary would be composed of only \textsf{poly}$(\log d)$ gates, where $d$ is the  Hilbert space in which the unitary resides. 

\section{\textsf{QPUF} from unitary $t$-designs} \label{sec:qpufdesign}

To overcome the issue of inefficient construction of \textsf{QPUF} as a unitary sampled uniformly randomly from Haar measure, we introduce a resource efficient construction of these devices using a unitary sampled uniformly randomly from the $t$-design set of unitaries. For a brief introduction into the designs, we refer the reader to section~\ref{sec:prel-design}. 

\subsection{$\textsf{QPUF}_t$ Generation and Evaluation}

Similar to the original proposal of \cite{arapinis2019quantum}, a $\textsf{QPUF}_t$ involves a manufacturing process where the manufacturer samples uniformly from set of unitaries $\mathcal{U} : \{\U_1,\cdots \U_k \} \in \U(d)$ acting on $n=\log_2 d$ qubits which form an approximate unitary $t$-design. This is formally defined via the manufacturer's quantum generation algorithm, 
\begin{equation}
    \textsf{QGen} : \textsf{QPUF}_{t, \textsf{id}} = U_{\textsf{id}} \stackrel{\$}{\leftarrow}  \{\U_1,\cdots \U_k \}
    \end{equation}

%The unique identifier corresponds to the specific unitary that is sampled in the \textsf{QGen} process. 
The $\textsf{QPUF}_t$ generated via this process needs to have good completeness requirements of robustness, collision-resistance and uniqueness. What could be potentially  problematic in the case of constructions of $\epsilon$-approximate unitary $t$-designs is the uniqueness property. The reason behind this is that these constructions usually \emph{sample} from a set of unitaries forming an approximate design \cite{BHH16}. Depending on the construction, this sampling (or equivalently the \textsf{QGen} algorithm) could be highly redundant in the sense that it samples the same unitary with high probability when implemented twice in succession. Furthermore, these identical unitaries could have different identifiers. For example, in the RQC model different runs of RQC's characterized by different two-qubit gates applied at different positions could (in this case, the unique identifier is the set of gates and set of positions where these gates are applied), depending on the two-qubit gate set used, produce the same overall unitary (possibly up to global phase).   This is detrimental to the uniqueness property which, as seen previously, demands that two QPUFS with two different identifiers be distinguishable with high probability.  To overcome this issue, we use a specific construction for $\varepsilon$-approximate $t$-designs which we will detail in the next paragraph.

Our \textsf{QGen} process involves a particular type of  circuit sampling from an approximate $t$-design inspired by the measurement based quantum computing (MBQC) approach \cite{RB01,MGDM18}, and from a recent result of \cite{OHS20}. 
%Here, we adapt the \emph{parallel random circuit} construction of \cite{OHS20} to the MBQC framework in order to arrive at a constant depth quantum circuit for sampling from a $\varepsilon$-approximate unitary $t$-design.
\footnote{We note that our construction can also be viewed in the MBQC framework as a 2D graph state which results in a constant depth quantum circuit \cite{mezher2020fault,bermejo2018architectures}}.
Note that our circuit is closely related to that of \cite{haferkamp2019closing}, with the main difference being that the circuit in \cite{haferkamp2019closing} was shown to sample from an approximate 2-design, whereas our construction samples from  an approximate $t$-design for any $t$, which is key for our security definition of \emph{practical unknownness} defined over the next sections. Although the results of \cite{haferkamp2019closing} should in principle be extendable to any $t>2$, it is not immediately clear from \cite{haferkamp2019closing} how this can be done, hence we do not consider their construction in this work. We now explain the paralell random circuit construction, a type of RQC on which our construction is based.

Let $G \subset \U(4)$ be a set of unitaries which is \emph{approximately universal} on $\U(4)$, in the sense that any $\U \in \U(4)$ can be approximated to an arbitrary accuracy $\gamma$ (in some norm)  by a product of  lenght $l(\gamma)$ of unitaries $\prod_{i=1,..,l(\gamma)}\V_i$, where each $\V_i \in G$. The parallel random circuit construction was first defined in \cite{BHH16} as follows on a 1D circuit of $n$ qubits initially in some state $|\psi\rangle$.
\begin{itemize}
    \item \textsf{Step 1}: On the 1D circuit, apply with a probability 0.5, either the unitary $$\U_{12} \otimes \U_{34}\cdots \otimes \U_{n-1 n} \hspace{2mm}(\text{for even $n$)}\hspace{2mm}/ \hspace{2mm} \U_{12} \otimes \U_{34}\cdots \otimes \U_{n-2 n-1} \hspace{2mm}(\text{for odd $n$})$$  or the unitary $$\U_{23} \otimes \U_{45}\cdots \otimes \U_{n-2 n-1} \hspace{2mm}(\text{for even $n$)}\hspace{2mm}/ \hspace{2mm} \U_{23} \otimes \U_{45}\cdots \otimes \U_{n-1 n} \hspace{2mm}(\text{for odd $n$})$$ where $U_{ii+1}$ for $i=1,..,n-1$ is a unitary chosen uniformly at random from $G$ and which acts on qubits $i$ and $i+1$ of the 1D circuit
    \item  \textsf{Step 2}: Repeat \textsf{Step 1} $k \geq C(G)t^9(2ntlog(2)+log(\dfrac{1}{\varepsilon}))$ times, where $t$ is a positive integer, $C(G)$ is a positive constant which depends on the choice of gate set $G$, and $0<\varepsilon<1$.
    
\end{itemize}
The output of this RQC is the state $\U_{\textsf{tot}}|\psi\rangle$, where $\U_{\textsf{tot}}$ is the product of all unitaries applied in the parallel random circuit construction, and which was shown in \cite{BHH16} to be a unitary sampled from a $2\varepsilon$-approximate unitary $t$-design. However, the main restriction in \cite{BHH16} was that the unitaries in $G$ had to be \emph{symmetric} ( if $\V \in G$, then $\V^{\dagger} \in G$) and composed entirely of algebraic entries, this strictly limits the choice of allowed gate sets $G$. However, the main result of \cite{OHS20} (and to an extent also that of \cite{MGDM20}) was removing the symmetric and algebraic entries requirements. Indeed, \cite{OHS20} show that $any$ choice of an approximately universal gate set $G \in \U(4)$ results in a $2\varepsilon$-approximate $t$-design in a parallel random circuit construction where $\textsf{Step 1}$ is repeated $\geq O(k)$  times, with $k$ as defined in $\textsf{Step 2}$.
Using the relaxation of \cite{OHS20}, we construct our parallel random circuit  where our gate set $G$ will be of the form $$G=\{CZ \cdot (X(\alpha+m_1\pi)Z(\beta+m_2\pi) \otimes (X(\gamma+m_3\pi)Z(\delta+m_4\pi))\}_{m_1,..,m_4,\alpha,\beta,\delta,\gamma},$$ where $\alpha, \beta, \gamma, \delta$ range over all angles in the interval $[0,2\pi]$, and $m_i \in \{0,1\}$ for $i=1,..4$.

\begin{figure}[h!]
\includegraphics[scale=0.5]{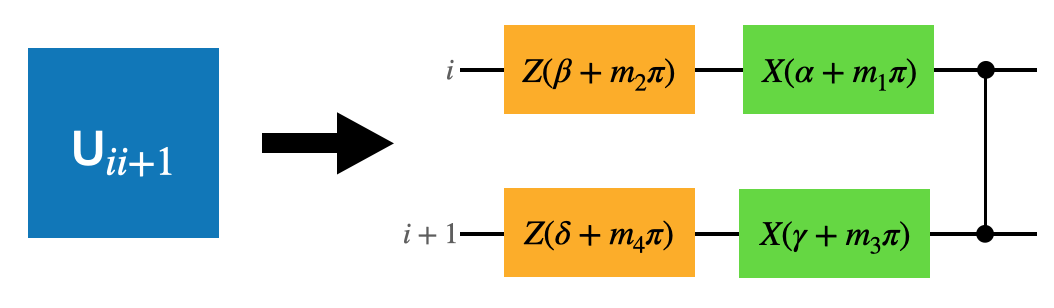}
\centering
\caption{An instance of the block unitary $\U_{ii+1}$ chosen uniformly from the set $G$. The unitary $\U_{ii+1}$ acts on the qubits $i$ and $i+1$, where $\alpha,\beta,\delta,\gamma$ are chosen uniformly from $[0,2\pi]$ and $m_1,m_2,m_3,m_4$ are bits chosen independently with probability 1/2 from the set $\{0,1\}$.}
\label{small-block}
\end{figure}

Thus, choosing $\U_{ii+1}$ (whose circuit implementation is shown in Figure ~\ref{small-block}) uniformly from $G$ corresponds to applying a unitary $$\U_{ii+1}=CZ_{ii+1}\cdot(X_i(\alpha+m_1\pi)Z_{i}(\beta+m_2\pi) \otimes X_{i+1}(\gamma+m_3\pi)Z_{i+1}(\delta+m_4\pi)),$$ acting on qubits $i$ and $i+1$ of the 1D circuit, where $\alpha,\beta,\delta,\gamma$ are chosen uniformly from $[0,2\pi]$ and $m_1,m_2,m_3,m_4$ are random bits chosen independently with probability 1/2 from the set $\{0,1\}$. 

The reason why the binary bits $m_i$ appear is because our construction is based upon an MBQC construction similar to those in \cite{MGDM19,haferkamp2019closing,MGDM18} which have been provably shown to converge to approximate $t$-designs. The bits $m_i$ have a natural interpretation in this setting as the random measurement results of single qubits in some basis (see for example \cite{RB01}). Although we will not go into details of MBQC constructions here, we keep this structure of the gate set to illustrate that one can go straightforwardly between both circuit and MBQC models using translations such as those in \cite{MGDM19}. This translation may be useful in implementing our construction on different hardware, such as, for example trapped ions \cite{lanyon2013measurement}.

Note that the choice of the angles $\alpha, \beta, \gamma, \delta$ uniformly from $[0,2\pi]$ will give us a good uniqueness. This is demonstrated by our numerical findings seen later on in section 4.3. The intuition behind why this should be true is that running two different instances of our construction, it would be highly unlikely that the same sets of angles (or even close sets of angles) are obtained; therefore the two output unitaries of these two runs should be sufficiently distinct (and therefore distinguishable) with high probability.

\begin{figure}[t]
\includegraphics[scale=0.5]{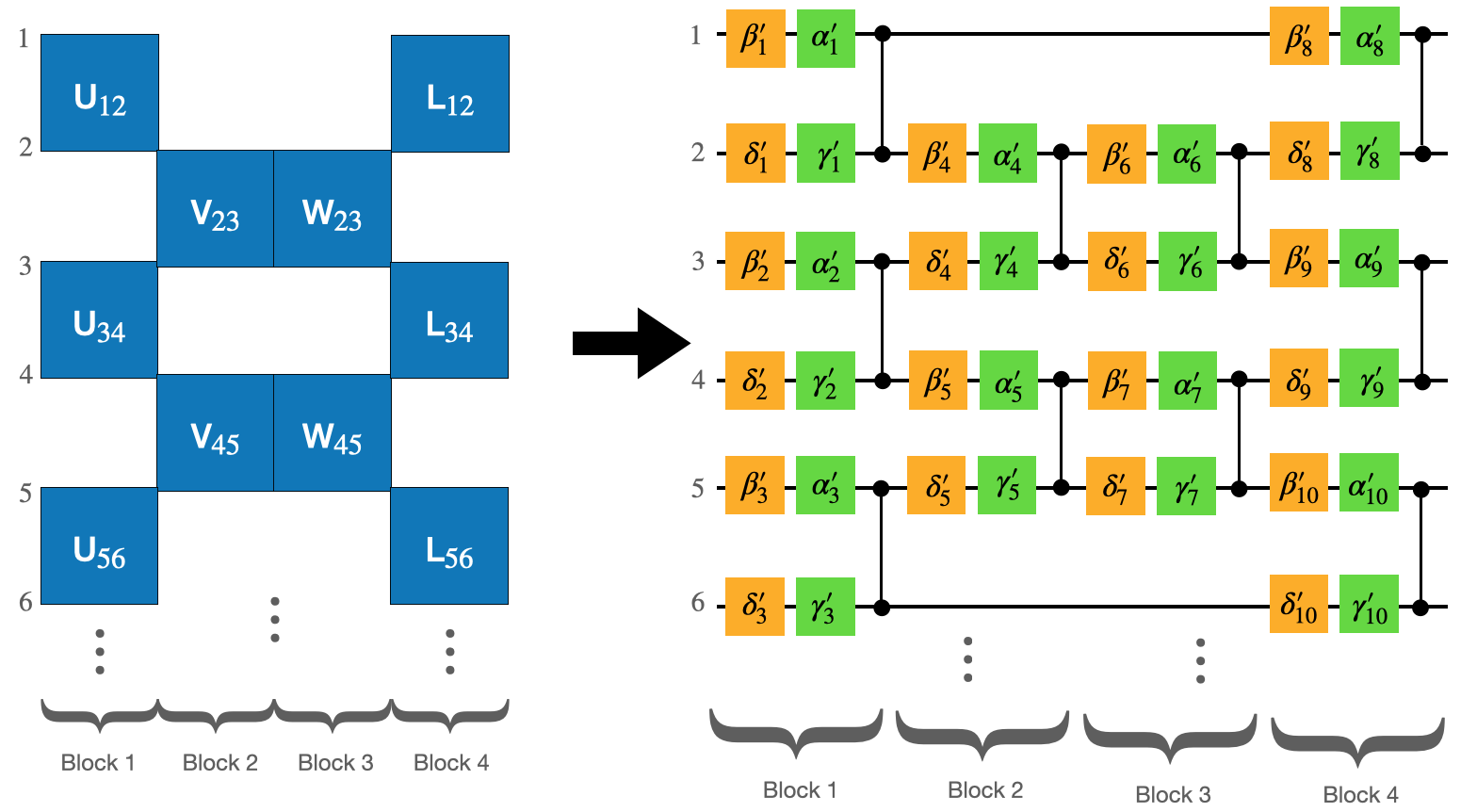}
\centering
\caption{An instance of a parallel local random circuit with $k=4$ (left), and its realisation in the circuit picture (right). Each block from 1 to 4 corresponds to \textsf{Step 1} in our parallel random circuit construction. The $U_{ii+1},V_{ii+1},W_{ii+1}$, and $L_{ii+1}$ are unitaries chosen uniformly at random from our chosen set $G$ and acting on qubits $i$ and $i+1$. As mentioned in the main text, our choice of $G$ and our parallel random circuit construction can be implemented naturally in the circuit picture with parameterised single qubit $X$ (in green) and $Z$ (in yellow) gates and the unparameterised CZ gate (right part of figure). The gate parameters $\alpha_i', \beta_i', \gamma_i', \delta_i'$ are equal to $\alpha_i + m_{1,i}\pi, \beta_i + m_{2,i}\pi, \gamma_{i} + m_{3,i}\pi, \delta_i + m_{4,i}\pi$ respectively, where $\alpha_i,\beta_i,\delta_i$, and $\gamma_i$ are $XY$ angles chosen uniformly from $[0,2\pi]$, and $m_{1,i}, m_{2,i}, m_{3,i}, m_{4,i}$ are bits chosen independently with probability 1/2 from the set $\{0,1\}$.}
\label{tdesqpuf}
\end{figure}

It can be easily seen that applying a parallel random circuit with $G$ as defined above amounts in the circuit picture to constructing the circuit composed of unitary blocks shown in Figure~\ref{small-block}\footnote{In the equivalent MBQC picture, this amounts to constructing a regular graph state followed by a series of $XY$  measurements where the angles are chosen from $[0,2\pi]$.}.  An instance of our parallel random circuit construction in the circuit picture is given in Figure \ref{tdesqpuf}.

We still need to prove that our chosen set $G$ is approximately universal on $\U(4)$, as required in the construction of approximate designs we use \cite{OHS20,BHH16}. We will use a result of \cite{brylinski2002universal} which shows that the set of all single qubit gates together with an entangling gate is approximately universal on $\U(4)$. Observe that the set $\{X(\alpha),Z(\beta)\}$ with $\beta$ and $\alpha$ ranging from 0 to $2\pi$ is dense in $\U(2)$. Indeed, any single qubit unitary can be represented as a product of gates from this set. Furthermore, $CZ$ is an entangling gate. Thus, our chosen set $G$ is approximately universal on $\U(4)$ from \cite{brylinski2002universal} since it contains a dense set in $\U(2)$ together with an entangling gate.

The evaluation process of this construction is captured by $\textsf{QEval}$ which maps the any input state $\rho_{in} \in \mathcal{H}^d$ to the output state $\rho_{out} \in \mathcal{H}^{d}$,
\begin{equation}
    \textsf{QEval} : \rho_{out} = \textsf{QPUF}_{t, \textsf{id}}\rho_{in}\textsf{QPUF}_{t, \textsf{id}}^{\dagger}
\end{equation}

Next we show that our construction satisfies the security notions characterised by robustness, collision-resistance, uniqueness and unknownness requirements. 

\subsection{Robustness $\&$ Collision-Resistance}
Following the same arguments as in section~\ref{sec:robust-haar}, the unitary property of $\textsf{QPUF}_t$ ensures that the device satisfies the robustness and collision-resistance properties. 

\subsection{Uniqueness}

In this section, we provide numerical evidences that our parallel random circuit based approximate $t$-design construction of $\textsf{QPUF}_t$ satisfies the uniqueness condition with an overwhelmingly high probability. For our numerical analysis, we construct two instances $\U_0$ and $\U_1$ of our parallel random circuit as shown in Figure~\ref{tdesqpuf}. This amounts to randomly choosing the parameters of the $X$ and $Z$ gates and computing the diamond norm between the resulting unitaries. We are able to infer that the diamond norm between any two such unitaries is overwhelmingly close to 2 (which corresponds to maximum distinguishability between the two unitaries). Further, we infer that the diamond norm closeness to 2 increases as one increases the circuit size : number of qubits and blocks. In the following figures, we plot the diamond norm between two unitaries picked using our construction (on the $y$-axis) with the number of instances (runs) of creating these unitaries (on the $x$-axis). The diamond norm in the plots between two such unitary channels is computed using the Equation~\ref{diamond_numerical_range} and Lemma~\ref{lemma:num-range-theta}. In other words, the diamond norm between $\U_0$ and $\U_1$ is expressed as a function of the numerical range $\delta (\U_0^{\dagger}\U_1)$. Further, this numerical range is computed using the convex hull of the eigenvalues of $\U_0^{\dagger}\U_1$. This in turn is computed using Lemma~\ref{lemma:num-range-theta} which tells us that the numerical range is equal to the minimum difference in angle (radians) of the two distinct eigenvalues of $\U_0^{\dagger}\U_1$. Note that the eigenvalue ($\alpha$) of a unitary matrix is of the form $e^{i\alpha}$, where $\alpha$ is the angle associated with the eigenvalue in the complex plane.  

\begin{figure}[h!]
\includegraphics[scale=0.75]{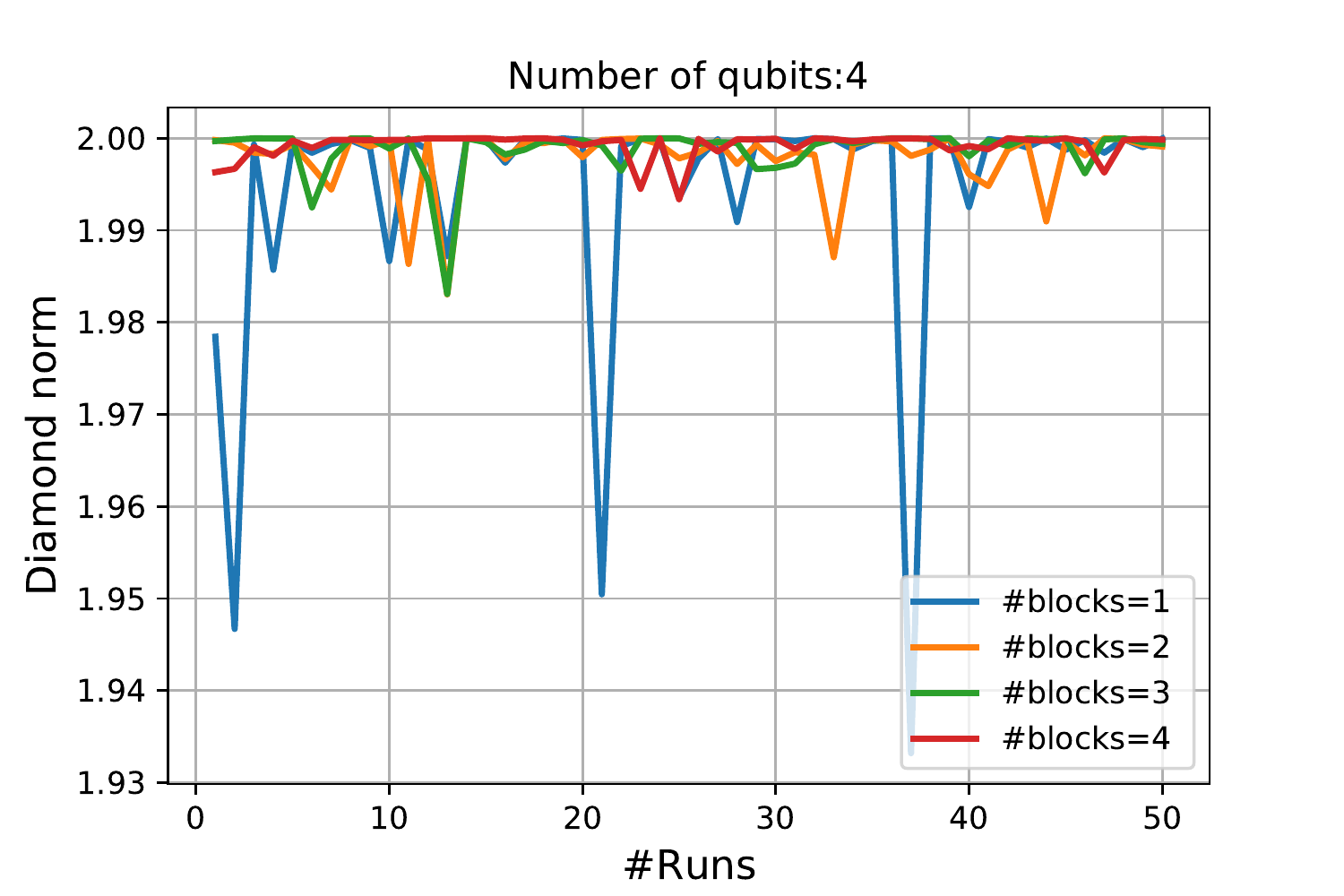}
\centering
\caption{Figure depicting the diamond norm vs number of runs for two independent circuits construction using our $\textsf{QPUF}_t$ construction for circuit of size corresponding to 4 qubits. We observe that across the 50 runs, the diamond norm for circuits corresponding to blocks 1, 2, 3, and 4 are lower bounded by 1.93, 1.98, 1.98, and 1.99 respectively. }
\label{fig:n-4}
\end{figure}

\begin{figure}[h!]
\includegraphics[scale=0.75]{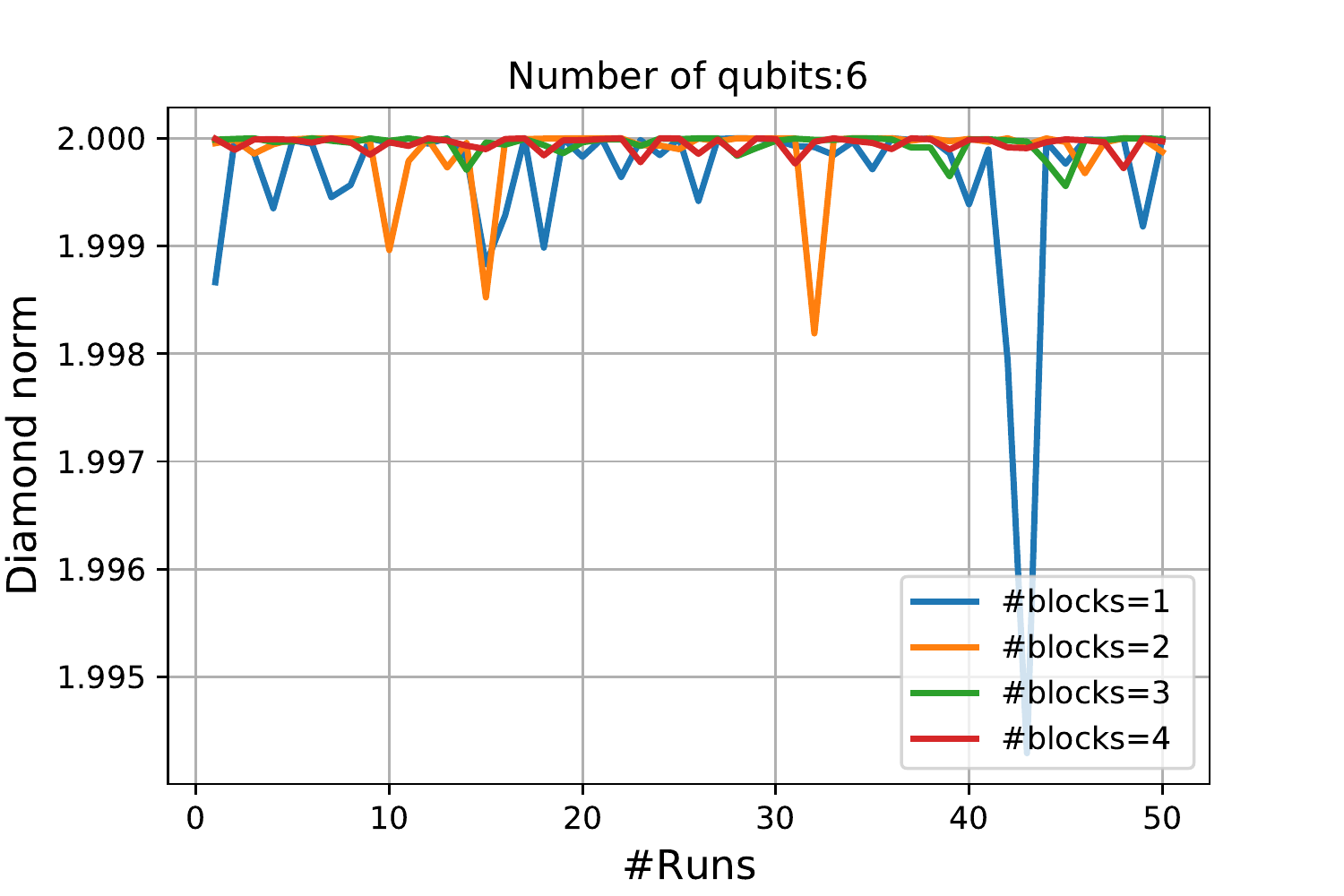}
\centering
\caption{Plot of diamond norm vs number of runs for circuit of size corresponding to 6 qubits. We observe that across the 50 runs, the diamond norm for circuits corresponding to blocks 1, 2, 3, and 4 are lower bounded by 1.994, 1.998, 1.999, and 1.999 respectively. }
\label{fig:n-6}
\end{figure}

\begin{figure}[h!]
\includegraphics[scale=0.75]{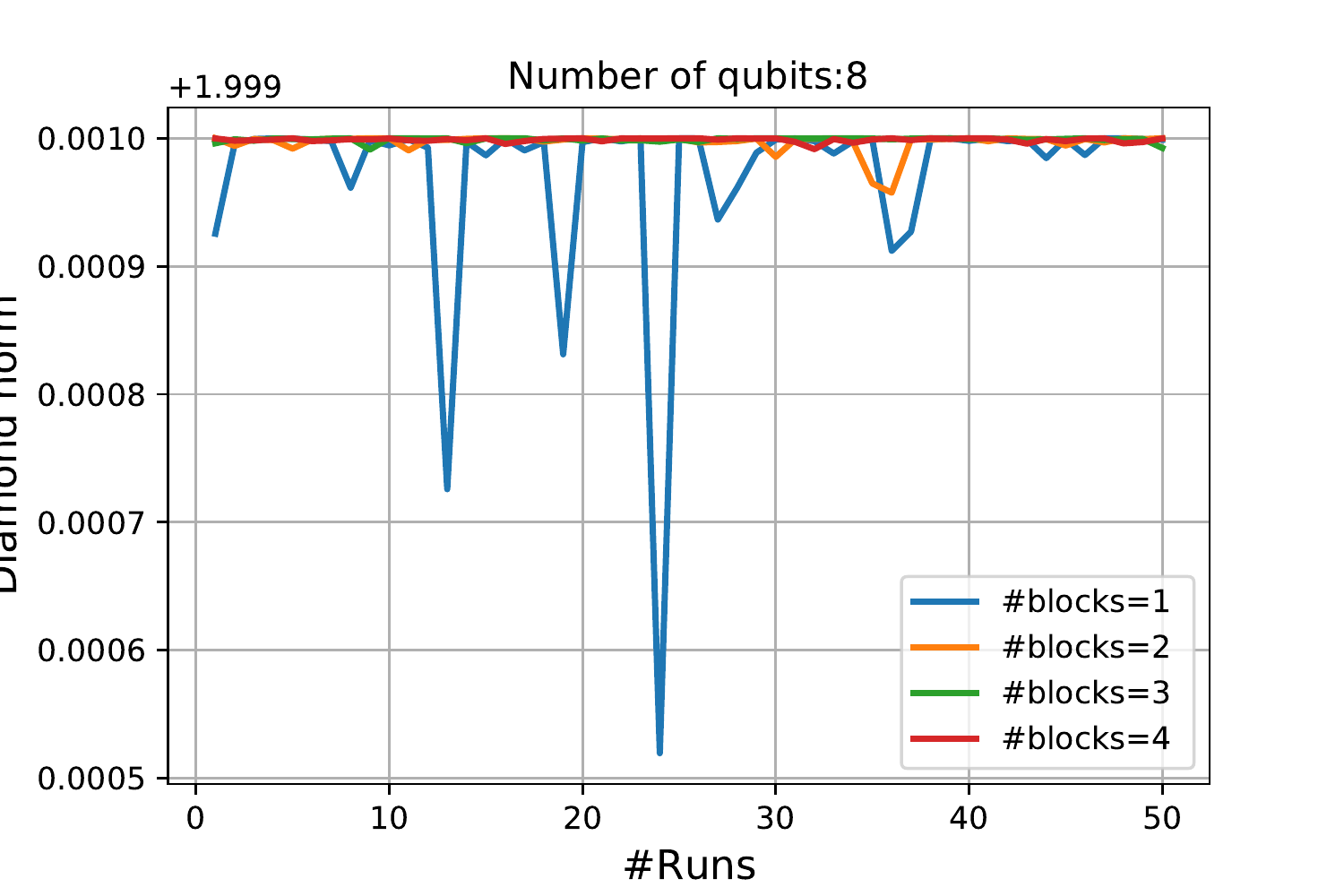}
\centering
\caption{Plot of diamond norm vs number of runs for circuit of size corresponding to 8 qubits. We observe that across the 50 runs, the diamond norm for circuits corresponding to blocks 1, 2, 3, and 4 are lower bounded by 1.99952, 1.99996, 1.99999, and 1.99999 respectively.}
\label{fig:n-8}
\end{figure}

Figures~\ref{fig:n-4},\ref{fig:n-6},\ref{fig:n-8} correspond to the plots of diamond norm vs number of runs for number of qubits ($\log_2 d$) corresponding to 4, 6 and 8 respectively. Each figure is plotted by considering up-to 4 blocks. Within each figure, we see that the closeness to the maximum value of 2 increases by adding more blocks in the circuit. This would be expected since adding more blocks amounts to randomly choosing parameters for more gates and hence the randomness in the circuit increases. Thus the probability that two such randomly picked circuits would be the same, decreases with increasing number of blocks. 

\subsection{Unknownness property}

As highlighted in the previous sections, the unknownness property of \textsf{QPUF} is essential to show the soundness property against an adversary trying to leverage information of the device in the query access model. We pointed out in section~\ref{sec:haar-qpuf-unknown} that the unknownness definition initially proposed for the Haar-random construction of \textsf{QPUF} ($\U \in \U(d)$) by \cite{arapinis2019quantum} implies that any QPT adversary, who has not apriori queried $\U$, has negligible probability of producing the correct response $\U\rho\U^{\dagger}$, for all input states $\rho \in \mathcal{H}^d$. We also mentioned that such an construction of a Haar random sampled unitary is impractical in the sense that the size of quantum circuits needed to sample from the Haar measure is exponential in $\log_2 d$. 

The above definition of unknownness does not allow constructing a \emph{practical} \textsf{QPUF} (i.e a \textsf{QPUF} generated by a $poly(\log_2 d)$-time algorithm). To see this, suppose that $\U$ can be generated by a quantum circuit $\mathcal{C}$ composed of $poly(\log_2 d)$ gates. Suppose $\mathcal{A}$  implements $\mathcal{C}$ (it is well within $\mathcal{A}$'s power to do so, since we assume a polynomial time adversary). $\mathcal{A}$ now has exactly reproduced $U$, and therefore can produce with probability one a correct response (i.e a response which is exactly equal to the ideal response $\U\rho\U^{\dagger}$), for all states $\rho \in \mathcal{H}^d$. The above definition rejects any such construction of a \textsf{QPUF} (since the probability of producing a correct response over all the states should be negligible for any $poly(\log_2 d)$ quantum circuit $\mathcal{C}$). Therefore, the only circuits allowed for generation of \textsf{QPUF} in the above definition are super-polynomial time circuits.

In order to get around this problem, we adopt a \emph{modified} definition of unknownness than that of \cite{arapinis2019quantum}, but which nevertheless is a reasonable definition of what a \textsf{QPUF} should be, and furthermore is satisfied naturally for $\varepsilon$-approximate unitary $t$-designs. We will call this \emph{practical} unknownness.
\begin{definition}[$\epsilon, t, d$-  \textbf{Practical unknownness}]: We say that a \textsf{QPUF} constructed as a unitary transformation $\U$ from a set $S \subseteq \U(d)$ is $(\epsilon,t,d)$- practically unknown if provided a bounded number $t \leq poly(\log_2 d)$ of queries $\U\rho\U^{\dagger}$, for any $\rho \in \mathcal{H}^d$,  the probability that any $poly(\log_2 d)$-time adversary can perfectly distinguish $\U$ from a Haar distributed unitary is upper bounded by $1/2(1 + 0.5\varepsilon)$.
\end{definition}
Here $0<\varepsilon<1$ and $t$ are functions of $\log_2 d$, and $lim_{log_2(d) \to \infty}\varepsilon=0$. 
\bigskip

Here distinguishability of two states $|\psi\rangle$ and $|\phi\rangle$ (for simplicity we describe the distinguishability in terms of pure states, but it can be naturally extended to mixed states) implies the ability to distinguish between $|\psi\rangle$  and $|\phi\rangle$ by performing some unitary transformation $\V$ on each of these states followed by measuring  the qubits of these two states in some basis. This definition of distinguishability is captured naturally in the definition of the $\|.\|_1$, a norm used to quantify the behaviour of an approximate $t$-design \cite{BHH16}. Indeed, when $\||\psi\rangle \langle \psi|-|\phi\rangle \langle \phi|\|_1=2$ it means that $|\phi\rangle$  and $|\psi\rangle$ are orthogonal, and thus maximally distinguishable; conversely when $\|\psi\rangle \langle \psi|-|\phi\rangle \langle \phi|\|_1=0$, then $|\psi\rangle=|\phi\rangle$ and these states are indistinguishable. Suppose we are given two states $\rho_1$ and $\rho_2$ uniformly at random, and we wish to derive a procedure for distinguishing $\rho_1$ from $\rho_2$. Any procedure we can think of amounts to performing a unitary $\V$ followed by some positive operator-valued measure (POVM), which could be 2-valued for a sophisticated enough choice of $\V$. Suppose we want to assign the measurement outcome 0 to $\rho_1$ and 1 to $\rho_2$ (in that case the POVM is 2-valued). In that case, it can be shown (see for example \cite{blume2015distinguishable}) that the probability of distinguishing $\rho_1$ from $\rho_2$ (i.e getting the measurement result 0 and having the state  $\rho_1$, or getting the measurement result 1 and having the state $\rho_2$) is given by

$$P_{distinguish}=\dfrac{1}{2}+\dfrac{1}{4}\|\rho_1-\rho_2\|_1.$$

Thus, if $\|\rho_1-\rho_2\|_1=2$ there is a procedure which can perfectly distinguish $\rho_1$ from $\rho_2$. On the other hand, if $\|\rho_1-\rho_2\|_1=0$ no process can do better than guess (with probability 1/2) whether the state is $\rho_1$ or $\rho_2$, and in that case the states are completely indistinguishable. In general, the smaller the norm $\|\rho_1-\rho_2\|_1$, the less distinguishable are  the states $\rho_1$ and $\rho_2$.

To see why this definition of unknownness is natural for unitary $t$-designs, consider a given a unitary $\U$ unknown to the adversary $\mathcal{A}$, and which is sampled from an $\varepsilon$-approximate $t$-design $\{p_i,\U_i\}_{i=1,\cdots, k}$. Suppose the adversary has access to $m \leq t$  copies of $\U\ket{\phi}$ , where $\ket{\phi} \in \mathcal{H}^d$ is any pure state known to $\mathcal{A}$, and possibly of his choosing. The mixed state seen by the adversary is then
\begin{equation}
  \rho=\sum_{i=1,\cdots, k}p_i\U_i^{\otimes m}(|\phi\rangle  \langle \phi|)^{\otimes m}\U_i^{\dagger, \otimes m}
\end{equation}
where the sum ranges over all elements of $\{p_i,\U_i\}_{i=1,\cdots, k}$. By the  definition of a $\varepsilon$-approximate $t$-design seen previously \cite{BHH16},
\begin{equation}
\label{eqtdes}
  \|\rho-\rho_H\|_1 \leq  sup_{l}\|\delta_{m} \otimes \mathbb{I}_l (X)-\delta^{H}_{m} \otimes \mathbb{I}_l (X) \|_1 \leq \|\delta_{m}-\delta^{H}_{m}\|_{\diamond} \leq \varepsilon,
\end{equation}
The $\|.\|_1$ norm, as seen before (see $P_{distinguish}$), relates directly to the distinguishability between two states, and is bounded in the case of approximate $t$-designs by an arbitrarily small chosen constant $\varepsilon$. Furthermore, in standard constructions of $t$-designs \cite{MGDM18,BHH16}, one can choose $t=poly(\log_2 d)$, and $\varepsilon=O(1/\exp(\log_2 d))$ with only a $poly(\log_2 d)$ increase in the depth of the quantum circuit sampling from the approximate $t$-design. Note that in this framework, we get a natural interpretation of the order $t$ of the design, it is simply the maximal number of queries an adversary can make such that the unitary $U$ remains practically unknown.

Finally, we note that the desired soundness property of selective unforgeability of our $\textsf{QPUF}_t$ construction against any bounded quantum adversary (see section~\ref{prelim-qpuf} for definition on selective unforgeability) follows naturally from the proof presented by \cite{arapinis2019quantum} for Haar random unitary \textsf{QPUF} construction. This follows from the fact that the failure of an adversary to produce the correct response of a selected challenge state (not selected by the adversary) is solely due to the fact that the challenge state is chosen from the Haar measure. The unknownness property developed here only excludes any attacks where the adversary can leverage the \emph{pre-protocol} information of the \textsf{QPUF} and thus would be able to trivially copy the device and thus be successful in producing the responses of any challenge state (even if the challenge is not prepared by the adversary).

\section{Noise-resilience of $\textsf{QPUF}_t$} \label{sec:robust}

Real-world implementation of $\textsf{QPUF}_t$ construction would involve the presence of undesired noise in the circuits. In this section, we consider the effects of a special type of noise, unitary noise, and show that our $t$-design based \textsf{QPUF} construction is resilient to this noise model. Unitary noise maps is a special case of general noise map which transforms an ideal unitary $\U$ to another $\U'$ representing the experimental imperfections. $\U'$ should still be \emph{close enough} to $\U$, this closeness is quantified in Equation (\ref{eqnoisebound}) below.%Additive-$\epsilon_t$ noise model further implies that the transformed unitary $\U'$ is at-most $\epsilon_t$ far away from the ideal unitary $\U$ in the diamond norm, i.e. $$\|\U - \U'\|_{\diamond} \leqslant \epsilon_t$$

Assume that ideal manufacturer has an approximate $t$-design set consisting of $\mathcal{U}=\{\U_1,\cdots,\U_k\}$ elements that are sampled with the probabilities $\{p_1,\cdots,p_k\}$ respectively. However, in reality, due to experimental implementations, the unitaries $\{\U_1,\cdots,\U_k\}$ are noisy. Under the unitary noise model, every unitary $\U_i$ is subjected to a unitary noise channel $\Lambda_i$. More precisely, for a given quantum state $\rho$,  $\Lambda_i \circ \U_i (\rho)=\U^{'}_i\rho \U_i^{'\dagger}$, where $\U^{'}_i$ the unitary implemented due to the unitary noise. 

Under this noise model, we immediately see that the requirements of robustness and collision-resistance are still preserved for $\textsf{QPUF}_t$ since these requirements only require the construction to be a unitary map. The requirements of uniqueness and practical unknownness leverage the $t$-design sampling property of the construction. As we show now, our chosen noise model still preserves the unitary sampling from the approximate $t$-design. We show this with  theorem~\ref{thm_noise}.
\begin{theorem} \label{thm_noise}
If the noiseless $\textsf{QPUF}_t$ corresponds to $\epsilon$-approximate $t$-design, then the $\epsilon_t$ additive unitary noise $\textsf{QPUF}_t$ corresponds to $\epsilon + \epsilon_t$-approximate $t$-design. 
\end{theorem}

\noindent \emph{Proof}: If the $\epsilon_t$-unitary noise map is denoted by $\Lambda_i$, then for all $\U_i \in \mathcal{U}$, we have

\begin{equation}
\label{eqnoisebound}
    \|ad_{\U^{\otimes t}_i}-ad^{\Lambda_i}_{\U^{\otimes t}_i}\|_{\diamond} \leq \varepsilon_t,
\end{equation}
where $ad_{\U^{\otimes t}_i}(.)=\U^{\otimes t}(.)\U^{ \dagger \otimes t}$, $ad^{\Lambda_i}_{\U^{\otimes t}_i}(.)$ is defined similarly to $ad_{\U^{\otimes t}_i}(.)$ but with each $U_i$ now acted upon by the noise channel $\Lambda_i$, $0<\varepsilon_t<1$  is the noise parameter, which is usually a function of $t$, it can be understood naturally as the overall noise strength acting on the tensor product  $U^{\otimes t}_i$, where each of the $t$ copies of $U_i$ is acted upon by $\Lambda_i$.

Now, since $\{p_i,\U_i\}_{i=1,\cdots,k}$ is a $\varepsilon$-approximate unitary $t$-design it satisfies (see Equation (\ref{eqtdes}) ) \cite{Low10}
\begin{equation}
    \|\delta_t-\delta^{H}_t\|_{\diamond} \leq \varepsilon,
\end{equation}
where $\delta_t(Y)=\sum_{i}p_i\U_i^{\otimes m}Y\U_i^{\dagger, \otimes m}=\sum_{i}p_i ad_{\U^{\otimes t}_i}(Y)$, $\delta^{H}_{m} (Y)=\int ad_{\U^{\otimes t}}(Y)d\mu(\U)$ are as defined previously in Equation (\ref{eqtdes}), and we define $\delta^{\Lambda}_t(Y)=\sum_{i}p_i ad^{\Lambda_i}_{\U^{\otimes t}_i}(Y)$ to be the noisy version of $\delta_t(Y)$ where each $\U_i$ is acted upon by the noise channel $\Lambda_i$.
By a triangle inequality,
$$\|\delta^{\Lambda}_t-\delta^{H}_t\|_{\diamond} \leq \|\delta^{\Lambda}_t-\delta_t\|_{\diamond}+\|\delta_t-\delta^{H}_t\|_{\diamond} \leq  \|\delta^{\Lambda}_t-\delta_t\|_{\diamond} + \varepsilon.$$  
Now,
$$\|\delta^{\Lambda}_t-\delta_t\|_{\diamond} \leq \sum_ip_i\|ad^{\Lambda_i}_{\U^{\otimes t}_i}-ad_{\U^{\otimes t}_i}\|_{\diamond} \leq \varepsilon_t\sum_ip_i \leq \varepsilon_t. $$
Replacing this in the above equation we obtain
\begin{equation}
    \label{eqrobusttdes}
    \|\delta^{\Lambda}_t-\delta^{H}_t\|_{\diamond} \leq \varepsilon + \varepsilon_t.
\end{equation}
Equation (\ref{eqrobusttdes}) is a meaningful definition of an $\varepsilon^{'}$-approximate $t$-design when $\varepsilon^{'}=\varepsilon+\varepsilon_t \leq 1$, this means that the noise strength should satisfy $\varepsilon_t \leq 1-\varepsilon$. Note that Equation (\ref{eqrobusttdes})  may not be a very practical definition of noise-resilience. Indeed it somehow demands that the overall noise strength $\varepsilon_t$ be at most a constant independent of $t$. This would mean, that as the system size increases, the error rates of single qubits should decrease so that this condition remains satisfied.  We expect that in order to get a better notion of noise-resilience, some notion of fault-tolerance and quantum error correction should be invoked.

\section{Discussion} \label{conclusion}

Quantum physical unclonable functions have become a rapidly emerging cryptographic technology to provide solutions for tasks including device and message authentication, secure key exchange among others. Although these have gained significant interest in recent times, a rigorous resource-efficient construction with provable security guarantees have been missing. Our work addresses this issue by proposing a \textsf{QPUF} based on an approximate unitary $t$-design construction; called $\textsf{QPUF}_t$. We utilise the parallel random circuit based $t$-design construction  \cite{BHH16,OHS20} and show that  this construction  provides provable security guarantees against a BQP adversary with bounded $t$-query access to the device. Further, our construction involves only nearest neighbour interaction of qubits, which makes our construction highly practical.
%thus making it tailor-made in the noisy-intermediate scale quantum regime. 
When viewed in the measurement based quantum computation model \cite{RB01,MGDM19,bermejo2018architectures}, our circuit is constant depth, which is a further indication of its potential near-term implementation. 

Our work also proposes the first use case of approximate $t$-designs for arbitrary $t$-values. We note that the previous use cases of $t$-designs have been restricted only to some (usually low) $t$ values; our work on the contrary states that higher the $t$ value our \textsf{QPUF} is constructed with, higher the level of security it provides. By this, we mean that we empower the adversary to be able to do more queries to the device in the pre-protocol in order to leverage key information of the device and thus use it during the protocol run. This can also be viewed alternatively that the \textsf{QPUF} can be allowed to remain in an unsecured environment for a longer period of time if its constructed with approximate designs of higher $t$ values (the adversary performs queries on the device while its in the unsecured environment). 

We realise that our work is just the beginning of (what could be) proposing efficient construction of these hardware-based cryptographic primitives with provable security guarantees. We envision the future work where these constructions are proven secure against adversaries who are not just restricted to the \emph{black-box} query access model. Specifically, security in the \emph{white-box} model would make these devices much more tailored to real world deployment. Under the white-box model, the adversary would know the circuit layout which would be in the public domain, but would not know the parameters of $X$ and $Z$ gates. Hence the security would depend on the classical randomness with which each of these parameters is chosen. 

Another natural extension of our work would be to prove noise-resilience for general noise maps (including stochastic noise maps). We believe that this would involve construction and the use of approximate $t$-design completely positive trace preserving (CPTP) channels instead of the approximate $t$-design unitaries that we consider in this work. This is due to the fact that a general noise map in the Kraus formulation is written as a general CPTP map. Hence it is important to prove that the noisy general CPTP \textsf{QPUF} would satisfy the required security notions.

\section{Acknowledgements}

We thank Mina Doosti, Mahshid Delavar and Myrto Arapinis for useful discussions. We acknowledge the support by the Engineering and Physical Sciences Research Council (grant EP/L01503X/1), Engineering and Physical Sciences Research Council Hub in Quantum Computing and Simulation EP/T001062/1 grant, and Verification of Quantum Technology (grant EP/N003829/1).

\medskip
\bibliographystyle{unsrt}
\bibliography{puf}

\begin{thebibliography}{10}

\bibitem{shannon1948mathematical}
Claude~E Shannon.
\newblock A mathematical theory of communication.
\newblock {\em The Bell system technical journal}, 27(3):379--423, 1948.

\bibitem{katz2020introduction}
Jonathan Katz and Yehuda Lindell.
\newblock {\em Introduction to modern cryptography}.
\newblock CRC press, 2020.

\bibitem{goldreich1998modern}
Oded Goldreich.
\newblock {\em Modern cryptography, probabilistic proofs and pseudorandomness},
  volume~17.
\newblock Springer Science \& Business Media, 1998.

\bibitem{mao2003modern}
Wenbo Mao.
\newblock {\em Modern cryptography: theory and practice}.
\newblock Pearson Education India, 2003.

\bibitem{canetti2001universally}
Ran Canetti and Marc Fischlin.
\newblock Universally composable commitments.
\newblock In {\em Annual International Cryptology Conference}, pages 19--40.
  Springer, 2001.

\bibitem{arapinis2019quantum}
Myrto Arapinis, Mahshid Delavar, Mina Doosti, and Elham Kashefi.
\newblock Quantum physical unclonable functions: Possibilities and
  impossibilities.
\newblock {\em arXiv preprint arXiv:1910.02126v1}, 2019.

\bibitem{vskoric2012quantum}
Boris {\v{S}}kori{\'c}.
\newblock Quantum readout of physical unclonable functions.
\newblock {\em International Journal of Quantum Information}, 10(01):1250001,
  2012.

\bibitem{young2016quantum}
Robert~James Young, Jonny Roberts, and Utz Roedig.
\newblock Quantum physical unclonable function, 2016.

\bibitem{nikolopoulos2017continuous}
Georgios~M Nikolopoulos and Eleni Diamanti.
\newblock Continuous-variable quantum authentication of physical unclonable
  keys.
\newblock {\em Scientific reports}, 7:46047, 2017.

\bibitem{gianfelici2020theoretical}
Giulio Gianfelici, Hermann Kampermann, and Dagmar Bru{\ss}.
\newblock Theoretical framework for physical unclonable functions, including
  quantum readout.
\newblock {\em Physical Review A}, 101(4):042337, 2020.

\bibitem{doosti2020client}
Mina Doosti, Niraj Kumar, Mahshid Delavar, and Elham Kashefi.
\newblock Client-server identification protocols with quantum puf.
\newblock {\em arXiv preprint arXiv:2006.04522}, 2020.

\bibitem{yao2016quantum}
Yao Yao, Ming Gao, Mo~Li, and Jian Zhang.
\newblock Quantum cloning attacks against puf-based quantum authentication
  systems.
\newblock {\em Quantum Information Processing}, 15(8):3311--3325, 2016.

\bibitem{vskoric2013security}
Boris {\v{S}}kori{\'c}, Allard~P Mosk, and Pepijn~WH Pinkse.
\newblock Security of quantum-readout pufs against quadrature-based
  challenge-estimation attacks.
\newblock {\em International journal of quantum information}, 11(04):1350041,
  2013.

\bibitem{knill1995approximation}
Emanuel Knill.
\newblock Approximation by quantum circuits.
\newblock {\em arXiv preprint quant-ph/9508006}, 1995.

\bibitem{dankert2009exact}
Christoph Dankert, Richard Cleve, Joseph Emerson, and Etera Livine.
\newblock Exact and approximate unitary 2-designs and their application to
  fidelity estimation.
\newblock {\em Physical Review A}, 80(1):012304, 2009.

\bibitem{brandao2016local}
Fernando~GSL Brandao, Aram~W Harrow, and Micha{\l} Horodecki.
\newblock Local random quantum circuits are approximate polynomial-designs.
\newblock {\em Communications in Mathematical Physics}, 346(2):397--434, 2016.

\bibitem{epstein2014investigating}
Jeffrey~M Epstein, Andrew~W Cross, Easwar Magesan, and Jay~M Gambetta.
\newblock Investigating the limits of randomized benchmarking protocols.
\newblock {\em Physical Review A}, 89(6):062321, 2014.

\bibitem{hayden2004randomizing}
Patrick Hayden, Debbie Leung, Peter~W Shor, and Andreas Winter.
\newblock Randomizing quantum states: Constructions and applications.
\newblock {\em Communications in Mathematical Physics}, 250(2):371--391, 2004.

\bibitem{hayden2007black}
Patrick Hayden and John Preskill.
\newblock Black holes as mirrors: quantum information in random subsystems.
\newblock {\em Journal of high energy physics}, 2007(09):120, 2007.

\bibitem{hangleiter2018anticoncentration}
Dominik Hangleiter, Juan Bermejo-Vega, Martin Schwarz, and Jens Eisert.
\newblock Anticoncentration theorems for schemes showing a quantum speedup.
\newblock {\em Quantum}, 2:65, 2018.

\bibitem{harrow2018approximate}
Aram Harrow and Saeed Mehraban.
\newblock Approximate unitary $ t $-designs by short random quantum circuits
  using nearest-neighbor and long-range gates.
\newblock {\em arXiv preprint arXiv:1809.06957}, 2018.

\bibitem{bermejo2018architectures}
Juan Bermejo-Vega, Dominik Hangleiter, Martin Schwarz, Robert Raussendorf, and
  Jens Eisert.
\newblock Architectures for quantum simulation showing a quantum speedup.
\newblock {\em Physical Review X}, 8(2):021010, 2018.

\bibitem{BHH16}
Fernando~GSL Brandao, Aram~W Harrow, and Micha{\l} Horodecki.
\newblock Local random quantum circuits are approximate polynomial-designs.
\newblock {\em Communications in Mathematical Physics}, 346(2):397--434, 2016.

\bibitem{OHS20}
Micha{\l} Oszmaniec, Adam Sawicki, and Micha{\l} Horodecki.
\newblock Epsilon-nets, unitary designs and random quantum circuits.
\newblock {\em arXiv preprint arXiv:2007.10885}, 2020.

\bibitem{RB01}
Robert Raussendorf and Hans~J Briegel.
\newblock A one-way quantum computer.
\newblock {\em Physical Review Letters}, 86(22):5188, 2001.

\bibitem{mezher2020fault}
Rawad Mezher, Joe Ghalbouni, Joseph Dgheim, and Damian Markham.
\newblock Fault-tolerant quantum speedup from constant depth quantum circuits.
\newblock {\em arXiv preprint arXiv:2005.11539}, 2020.

\bibitem{preskill2018quantum}
John Preskill.
\newblock Quantum computing in the nisq era and beyond.
\newblock {\em Quantum}, 2:79, 2018.

\bibitem{armknecht2016towards}
Frederik Armknecht, Daisuke Moriyama, Ahmad-Reza Sadeghi, and Moti Yung.
\newblock Towards a unified security model for physically unclonable functions.
\newblock In {\em Cryptographers’ Track at the RSA Conference}, pages
  271--287. Springer, 2016.

\bibitem{DCE+09}
Christoph Dankert, Richard Cleve, Joseph Emerson, and Etera Livine.
\newblock Exact and approximate unitary 2-designs and their application to
  fidelity estimation.
\newblock {\em Physical Review A}, 80(1):012304, 2009.

\bibitem{B19}
Eiichi Bannai, Mikio Nakahara, Da~Zhao, and Yan Zhu.
\newblock On the explicit constructions of certain unitary t-designs.
\newblock {\em Journal of Physics A: Mathematical and Theoretical},
  52(49):495301, 2019.

\bibitem{BN20}
Eiichi Bannai, Yoshifumi Nakata, Takayuki Okuda, and Da~Zhao.
\newblock Explicit construction of exact unitary designs.
\newblock {\em arXiv preprint arXiv:2009.11170}, 2020.

\bibitem{HM18}
Aram Harrow and Saeed Mehraban.
\newblock Approximate unitary $ t $-designs by short random quantum circuits
  using nearest-neighbor and long-range gates.
\newblock {\em arXiv preprint arXiv:1809.06957}, 2018.

\bibitem{wieand2002eigenvalue}
Kelly Wieand.
\newblock Eigenvalue distributions of random unitary matrices.
\newblock {\em Probability Theory and Related Fields}, 123(2):202--224, 2002.

\bibitem{MGDM18}
Rawad Mezher, Joe Ghalbouni, Joseph Dgheim, and Damian Markham.
\newblock Efficient quantum pseudorandomness with simple graph states.
\newblock {\em Physical Review A}, 97(2):022333, 2018.

\bibitem{haferkamp2019closing}
Jonas Haferkamp, Dominik Hangleiter, Adam Bouland, Bill Fefferman, Jens Eisert,
  and Juani Bermejo-Vega.
\newblock Closing gaps of a quantum advantage with short-time hamiltonian
  dynamics.
\newblock {\em arXiv preprint arXiv:1908.08069}, 2019.

\bibitem{MGDM20}
Rawad Mezher, Joe Ghalbouni, Joseph Dgheim, and Damian Markham.
\newblock On unitary t-designs from relaxed seeds.
\newblock {\em Entropy}, 22(1):92, 2020.

\bibitem{MGDM19}
Rawad Mezher, Joe Ghalbouni, Joseph Dgheim, and Damian Markham.
\newblock Efficient approximate unitary t-designs from partially invertible
  universal sets and their application to quantum speedup.
\newblock {\em arXiv preprint arXiv:1905.01504}, 2019.

\bibitem{lanyon2013measurement}
BP~Lanyon, P~Jurcevic, M~Zwerger, C~Hempel, EA~Martinez, W~D{\"u}r, HJ~Briegel,
  Rainer Blatt, and Christian~F Roos.
\newblock Measurement-based quantum computation with trapped ions.
\newblock {\em Physical review letters}, 111(21):210501, 2013.

\bibitem{brylinski2002universal}
Jean-Luc Brylinski and Ranee Brylinski.
\newblock Universal quantum gates.
\newblock {\em Mathematics of quantum computation}, 79, 2002.

\bibitem{blume2015distinguishable}
Robin~J Blume-Kohout.
\newblock How distinguishable are two quantum processes?.
\newblock Technical report, Sandia National Lab.(SNL-NM), Albuquerque, NM
  (United States), 2015.

\bibitem{Low10}
Richard~A Low.
\newblock Pseudo-randomness and learning in quantum computation.
\newblock {\em arXiv preprint arXiv:1006.5227}, 2010.

\end{thebibliography}

\end{document}